\documentclass[prb,aps,twocolumn,showpacs,superscriptaddress]{revtex4}
\usepackage{graphicx}% Include figure files
\usepackage{epstopdf}
\usepackage{dcolumn}% Align table columns on decimal point
\usepackage{bm}% bold maths
\usepackage{longtable}
\usepackage{graphics}
\usepackage{amssymb}
\usepackage{amsmath}
\usepackage{xspace}
\usepackage{epsfig}

\begin{document}
\def \Cl{$\kappa$-(ET)$_2$Cu[N(CN)$_2$]Cl~}
\def \Cln{$\kappa$-(ET)$_2$Cu[N(CN)$_2$]Cl}
\def \CN{$\kappa$-(ET)$_2$Cu$_2$(CN)$_3$~}
\def \CNn{$\kappa$-(ET)$_2$Cu$_2$(CN)$_3$}
\def \Br{$\kappa$-(ET)$_2$Cu[N(CN)$_2$]Br~}
\def \Brn{$\kappa$-(ET)$_2$Cu[N(CN)$_2$]Br}
\def \NCS{$\kappa$-(ET)$_2$Cu(NCS)$_2$~}
\def \NCSn{$\kappa$-(ET)$_2$Cu(NCS)$_2$}
\def \cation{$\kappa$-(ET)$_2$}~
\def \kpx{$\kappa$-(ET)$_2X$}
\def \kbr{$\kappa$-(ET)$_2$Cu[N(CN)$_2$]Br}
\def \dBr{$\kappa$-(d8)-(ET)$_2$Cu[N(CN)$_2$]Br }
\def \dBrn{$\kappa$-(d8)-(ET)$_2$Cu[N(CN)$_2$]Br}
\def \hBr{$\kappa$-(h8)-(ET)$_2$Cu[N(CN)$_2$]Br }
\def \hBrn{$\kappa$-(h8)-(ET)$_2$Cu[N(CN)$_2$]Br}
\def \kcl{$\kappa$-(ET)$_2$Cu[N(CN)$_2$]Cl}
\def \kncs{$\kappa$-(ET)$_2$Cu(NCS)$_2$}
\def \kcn3{$\kappa$-(ET)$_2$Cu$_2$(CN)$_3$}
\def \>{\textgreater}
\def \<{\textless}
\def \q{\vec{q}}
\def \Q{\vec{Q}}
\def \kpcl{$\kappa$-Cl}
\def \kpbr{$\kappa$-Br}
\def \kpncs{$\kappa$-NCS}
\def \kpcn3{$\kappa$-(CN)$_3$}
\def \d8pbr{$\kappa$(d8)-Br}
\def \m{\mathrm{m}}
\def \max{\mathrm{max}}
\def \cross{\mathrm{cross}}
\def \M{\mathrm{M}}
\def \c{\mathrm{c}}
\def \lw{\mathrm{LW}}
\def \af{\mathrm{AF}}
\def \fm{\mathrm{FM}}
\def \sf{\mathrm{SF}}
\def \res{{\rho \propto T^2}}
\def \us{{\Delta v/v}}
\def \nmr{\mathrm{NMR}}
\def \coh{\mathrm{coh}}
\def \ks{{K_s}}
\def \exp{\mathrm{exp}}
\def \chiqw{$\chi({\bf q},\omega)$}
\newcommand {\ibid}{{\it ibid}. }
\newcommand {\etal}{{\it et al}. }
\newcommand {\etaln}{{\it et al}.}
\newcommand {\etalc}{{\it et al}., }

\title{Spin Fluctuations and the Pseudogap in Organic Superconductors}
\author{B. J. Powell}
\email{bjpowell@gmail.com}
\affiliation{Centre for Organic Photonics and Electronics, The University of Queensland, Brisbane,
Queensland 4072, Australia}

\author{Eddy Yusuf}
\affiliation{Centre for Organic Photonics and Electronics, The University of Queensland, Brisbane,
Queensland 4072, Australia}
\affiliation{Physics Department, University at
Buffalo, SUNY, Buffalo, NY 14260, USA}

\author{Ross H. McKenzie}
\affiliation{Centre for Organic Photonics and Electronics, The University of Queensland, Brisbane,
Queensland 4072, Australia}
\date{\today}

\begin{abstract}
We show that there are strong similarities in the spin lattice
relaxation of non-magnetic organic charge transfer salts, and that
these similarities can be understood in terms of spin fluctuations.
Further, we show that, in all of the $\kappa$-phase organic superconductors for
which there is nuclear magnetic resonance data, the energy scale for
the spin fluctuations coincides with the energy scale for the
pseudogap. This suggests that the pseudogap is caused by short-range
spin correlations. In the weakly frustrated metals $\kappa$-(BEDT-TTF)$_2$Cu[N(CN)$_2$]Br, $\kappa$-(BEDT-TTF)$_2$Cu(NCS)$_2$, and
$\kappa$-(BEDT-TTF)$_2$Cu[N(CN)$_2$]Cl (under pressure) the pseudogap opens at the same temperature as
coherence emerges in the (intralayer) transport. We argue that this
is because the spin correlations  are cut off by the loss of
intralayer coherence at high temperatures. We discuss what might
happen to these two energy scales at high pressures, where the
electronic correlations are weaker. In these weakly frustrated
materials the data is well described by the chemical pressure
hypothesis (that anion substitution is equivalent to hydrostatic
pressure). However, we find important differences in the metallic
state of  $\kappa$-(BEDT-TTF)$_2$Cu$_2$(CN)$_3$, which is highly frustrated and displays a spin
liquid insulating phase. We also show that the characteristic
temperature scale of the spin fluctuations in (TMTSF)$_2$ClO$_4$ is
the same as superconducting critical temperature, which may be
evidence that spin fluctuations mediate the superconductivity in the
Bechgaard salts.
\end{abstract}

\maketitle

\section{Introduction}

Strongly correlated superconductors such as the cuprates,\cite{LNW} heavy
fermions,\cite{stewart} and organic charge transfer salts,\cite{JPCMreview} %strontium ruthenate\cite{MeanoMackenzie} and sodium cobaltate,\cite{???}
share many phenomena not found in weakly correlated metals and
superconductors. Examples of such effects include, superconductivity
in close proximity to the Mott transition,\cite{LNW,JPCMreview}
metallic states not described by Landau's theory of Fermi
liquids,\cite{LNW,JPCMreview} small superfluid stiffnesses in the
superconducting state\cite{uemura,Tc-penetration} and pseudogap
phenomena.\cite{LNW,sidorov,Eddy1,Ong,Nam} Important questions about
these materials include: do these phenomena have a common origin and
how similar are the phenomena observed in the different families of
materials?

One common feature of the materials discussed above is that  they
display strong antiferromagnetic spin
fluctuations.\cite{Moriya,MMP,Eddy1,Curro-review} This can be seen
in the spin lattice relaxation rate, $1/T_1$, which is very
different from those found in weakly correlated materials (where $1/T_1\propto T$ cf. Ref \onlinecite{Eddy2}). In many cuprates, heavy fermion materials, and organic superconductors $1/T_1T$
is strongly temperature dependent: as the temperature is lowered
from room temperature $1/T_1T$ increases until it reaches a maximum
at a temperature we label $T_\nmr$. Below $T_\nmr$ a suppression of
spectral weight is manifest in $1/T_1T$ due to the opening of a
superconducting gap or a pseudogap. Throughout this paper, when we
refer to the pseudogap regime of the organic superconductors we mean
to indicate the temperatures between $T_\nmr$ and $T_c$, the
superconducting critical temperature, where there is a loss of
spectral weight evident in the NMR, but no bulk superconductivity.

It has  recently been shown that in the high temperature regime
(i.e., where there is neither a pseudogap or superconductivity)
simple scaling relations, based on a phenomenological two-fluid
picture, describe the spin lattice relaxation rate, $1/T_1$, in a
large number of high temperature
superconductors\cite{Barzykin,CurroMRS} and heavy fermion
materials.\cite{NPF,CYSP,Yang,CurroMRS} Further, it has been shown
that this two fluid model can describe many other experiments on the
heavy fermion materials.\cite{Yang,Yang-nature} Therefore, given the
similarities between these materials and the organic charge transfer
salts,\cite{Ross-science,JPCMreview} it is natural to ask whether
similar scaling behaviours describe the behaviour of the organic
charge transfer salts. In this paper we focus on this question in
the context of the $\kappa$-(ET)$_2X$ family of organic charge
transfer salts, however, we also make some comments on
(TMTSF)$_2$ClO$_4$.

%Two paradigms have emerged in which to consider superconductivity and pseudogap phenomena in unconventional superconductors. One is centred on the fact that these are strongly correlated materials and ascribes a central role to Mott physics - this approach contains theories such as the resonating valence bond theory.\cite{LNW,AndersonRVB,organicsRVB} The alternative, weak coupling approach, treats spin fluctuations as gauge bosons analogous to phonons in the BCS theory.\cite{MPL} Both of these approaches have been studied in the organics,\cite{JPCMreview} and understanding the behaviour of the spin degrees of freedom in the normal state is therefore important as it may provide evidence to differentiate between the two theories.

\begin{figure}%[h!t]
\epsfig{file=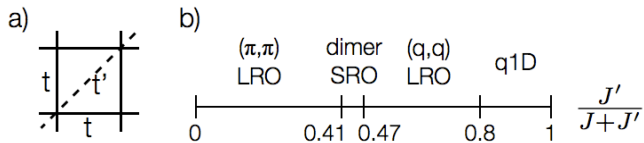,width=8.5cm}
\caption{(a) The anisotropic triangular lattice is believed to provide the basic description of the electronic structure of the $\kappa$-(ET)$_2X$ salts. This model has a tight binding structure where each site represents a dimer, (ET)$_2$. There is a hopping integral, $t$, along the sides of a square and another, $t'$, along one diagonal. Further, there is a strong Coulomb repulsion, $U$, if two electrons are places on the same site. For a review see Ref. \onlinecite{JPCMreview}. For $X$=Cu$_2$(CN)$_3$ $t'\simeq t$ and hence, in the Mott insulating phase, $J\simeq J'$, as $J\simeq t^2/U$ and $J'\simeq t'^2/U$. For the other $X$ discussed in this paper $t>t'$ and thus the geometrical frustration is significantly reduced. (b) Phase diagram of the Heisenberg model of the anisotropic triangular lattice from series expansion calculations,\cite{Weihong} which shows the sensitivity of this model to variations in $J'/J$. The following abbreviations are used in the figure: LRO (long range order), SRO (short range order), and q1D (qausi-one dimensional). ($\pi$,$\pi$) is the wavevector associated with N\'eel order and ($q$,$q$) is the wavevector for spiral ordering, which varies continuously from $q=\pi$ to $q=\pi/2$ as $J'/J$ increases.}
\label{fig:aniso}
\end{figure}

An important difference between the cuprates and heavy fermion
materials, and the $\kappa$-(ET)$_2X$ salts is that the
$\kappa$-phase organics there is a single half-filled band, whereas
the other materials are multiband and/or doped. The Mott transition
in $\kappa$-(ET)$_2X$, where $X$ is a monovalent anion, can be
driven by applying a hydrostatic pressure or varying the anion,
which is often referred to as chemical pressure. It is believed that
this corresponds to increasing the ratio of $t/U$ in the single band
Hubbard model description of these systems (cf. Fig. \ref{fig:aniso}).\cite{JPCMreview,Edan}

The behaviour of the spin degrees of freedom in  \CN are particularly interesting. Whereas the insulating phases of other
$\kappa$-(ET)$_2X$ salts order antiferromagnetically, the
insulating phase of \CN shows no signs of magnetic order to the
lowest temperatures investigated ($\sim$32 mK).\cite{Shimizu} As the exchange
energy extracted\cite{Shimizu,Zheng} from  fits to the high
temperature bulk susceptibility is $\sim$250 K this has been taken
as evidence that \CN is a spin liquid.\cite{Shimizu,JPCMreview} \CN
suffers from stronger geometrical frustration than other $\kappa$-(ET)$_2X$ salts: tight binding calculations\cite{Komatsu} indicate that the
band structure of \CN is close to that of an equilateral triangular lattice (i.e., $t\simeq t'$; cf. Fig. \ref{fig:aniso}), whereas
other $\kappa$-(ET)$_2X$ materials are described by an anisotropic triangular
lattice. ($t>t'$);\cite{JPCMreview} More importantly, fits
of high temperature series expansions to the bulk
susceptibility show that $J\simeq J'$ in \CN but that $J>J'$ in \Cln.\cite{Shimizu,Zheng} A pressure of greater than
$\sim$0.3 GPa drives the ground state of \CN from a spin liquid to a
superconductor. To date very little is known experimentally about
this superconducting state. It has recently been
argued\cite{group-theory,d+id} that the frustration will drive
changes in the spin fluctuations which will lead to a superconducting
state with broken time reversal symmetry, i.e., a $d+id$ (or more
strictly $A_{1}+iA_{2}$) state. Therefore it is important to understand the nature of the spin fluctuations in \CN in both the insulating and the metallic phases.

In order to investigate the spin fluctuations we present, below, a
theoretical analysis of previously published NMR experiments. In
section \ref{sect:highT} we show that although NMR experiments on
the organic charge transfer salts are rather similar to those on the
cuprates and heavy fermion materials there are some important
differences. A detailed analysis shows that, in the
$\kappa$-phase organics, the pseudogap energy scale is set by the
spin fluctuations, while in (TMTSF)$_2$ClO$_4$ the superconducting
gap is the same size as the characteristic spin fluctuations. In
section \ref{sect:coh} we compare the pseudogap energy scale with
the temperature at which coherence emerges in the intralayer
transport. We find that they are the same to within experimental
error and propose a phenomenological interpretation of this. In
section \ref{sect:highP} we discuss the behaviour of the organic
charge transfer salts under hydrostatic pressures and raise some
important issues about the pseudogap that have not yet been
addressed experimentally. Finally, we draw our conclusions in
section \ref{sect:conc}.

\section{High temperature spin lattice relaxation}\label{sect:highT}

\subsection{Scaling and the two-fluid model}

In the two-fluid model proposed by Pines \etal\cite{Barzykin,CurroMRS,NPF,CYSP,Yang} the spin lattice relaxation rate is given by
\begin{equation}
\frac{T_1T}{(T_1T)_\nmr}=\phi+\kappa T_\nmr\frac{T}{T_\nmr},
\end{equation}
where $\kappa$ and $\phi$ are material dependent constants, the latter measuring the proximity to a putative quantum critical point, and $T_\nmr$ is the temperature where there is a maximum in $1/T_1T$. The two fluids are proposed to be a Fermi liquid component and a spin liquid component.

\begin{figure*}%[h!t]
\epsfig{file=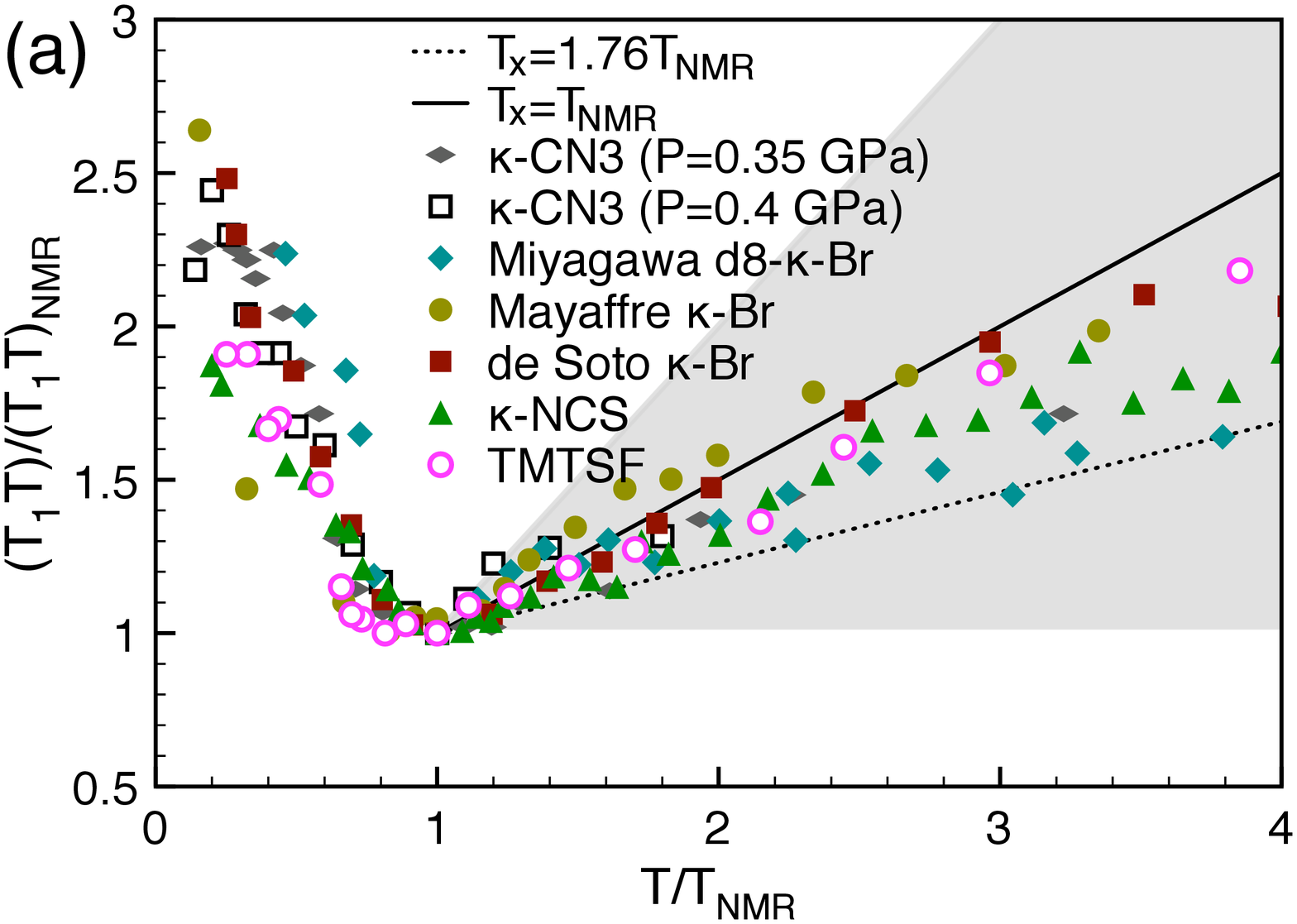,width=8cm}
\epsfig{file=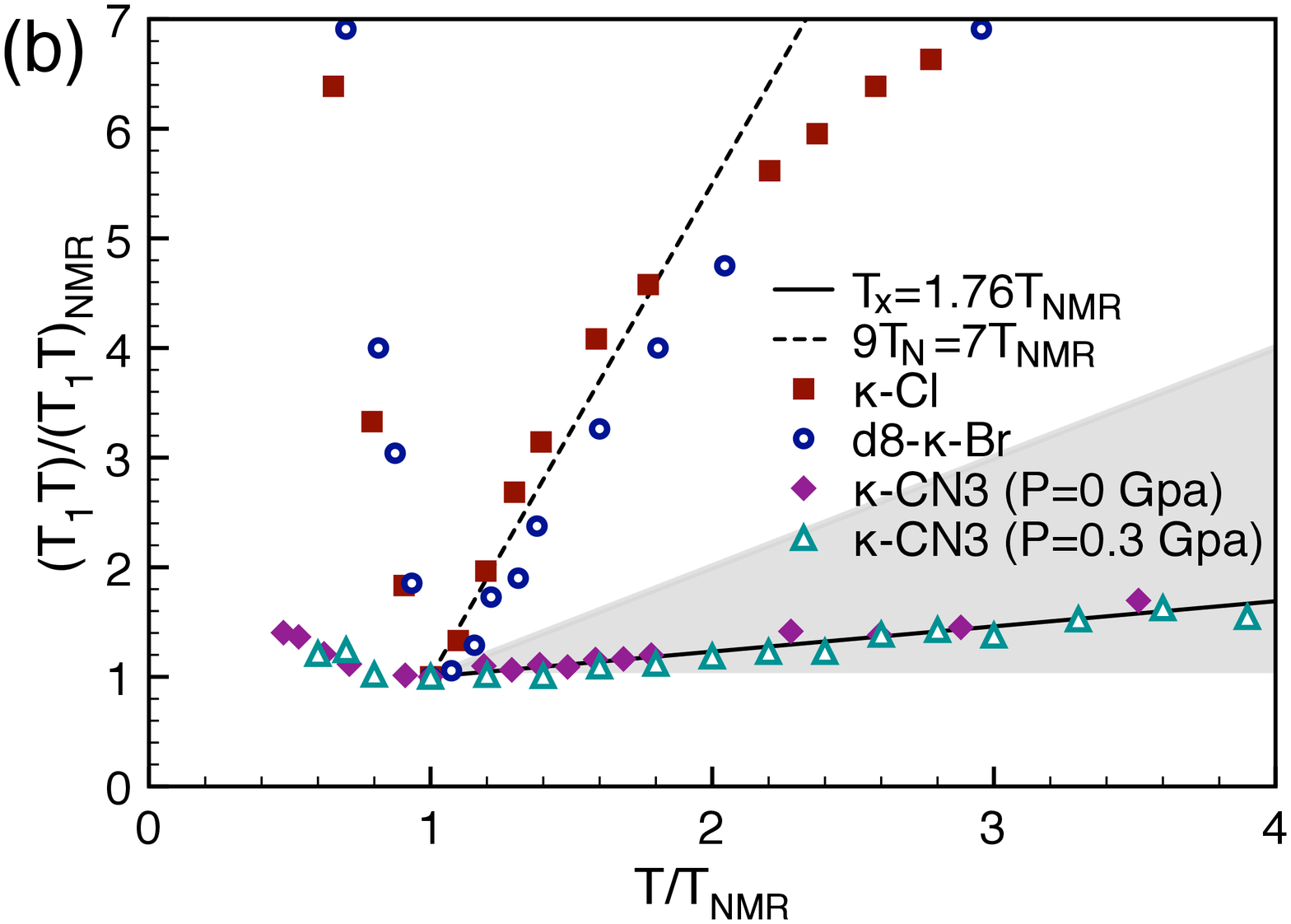,width=8cm}
\caption{(Color online)
Rescaled plots of the temperature dependence of $1/T_{1}T$ for  (a) metallic and  (b) insulating organic charge transfer salts. While there is a clear trend in the data of the non-magnetic materials, they do not collapse onto a single curve as the data for the cuprates and heavy fermion materials do.\cite{Barzykin,CurroMRS,NPF,CYSP,Yang} This suggests that a two-fluid fluid description is not required for the organic superconductors. However, this data is in good agreement with the prediction of the spin fluctuation model [Eqs. (\ref{scaling}) and (\ref{scalingAFM})]. Eq. (\ref{scaling}) predicts that the data in the will lie in the grey shaded regions, which represent the extrema of possible values of $T_x/T_\nmr$. The lines show the predictions for particular values of $T_x/T_\nmr$ in Eq. (\ref{scaling}) or $T_N/T_\nmr$ in Eq. (\ref{scalingAFM}) as marked. The abbreviations used in the figure and the sources of the data are, $\kappa$-CN3 is \CN (Ref. \onlinecite{kawamoto:kcn32006}), $\kappa$-Br is \Br (Refs. \onlinecite{deSoto,mayaffre}), d8-$\kappa$-Br is \dBr (Ref. \onlinecite{miyagawa-d-br}), $\kappa$-NCS is \NCS (Ref. \onlinecite{kawamoto}), TMTSF is (TMTSF)$_2$ClO$_4$ (Ref. \onlinecite{shinagawa}), and $\kappa$-Cl is \Cl (Ref. \onlinecite{miyagawa-kCl}).}
\label{fig:scaling}
\end{figure*}

We plot previously published\cite{kawamoto:kcn32006,mayaffre,deSoto,miyagawa-d-br,kawamoto,shinagawa,miyagawa-kCl} NMR data for both insulating and metallic phases of organic charge transfer salts in this `scaling' form in Fig. \ref{fig:scaling}. For the metallic salts it can be seen that, while there is a broad trend in the data, the data do not collapse onto a single curve as they do in the cuprates and the heavy fermion materials.\cite{Barzykin,CurroMRS,NPF,CYSP,Yang}
This suggests that two fluid model may not be relevant to the organic charge transfer salts, but also that there is something to be learnt from Fig. \ref{fig:scaling}.
For the insulating states the spin lattice relaxation is very different for those compounds that order magnetically (\Cl and fully dueterated \Br \{henceforth \dBrn\}) from that in the spin liquid \CNn. Indeed the data for \CN is remarkably similar to the data for the metallic salts.

Important evidence that a two-fluid model is required in the heavy fermion materials comes from comparing the Knight shift, $K_s$, to the bulk susceptibility $\chi$. According to the two fluid model
\begin{equation}
\chi=f(T)\chi_{FL}+[1-f(T)]\chi_{SL}, \label{eqn:2f-chi}
\end{equation}
where $f(T)$ is the fraction of electrons in the Fermi liquid, $\chi_{FL}$ is the susceptibility of the Fermi liquid, which is basically independent of temperature, and $\chi_{SL}$ is the susceptibility of the spin liquid. Further, the two-fluid model predicts that
\begin{equation}
K_s=Af(T)\chi_{FL}+B[1-f(T)]\chi_{SL}, \label{eqn:2f-Ks}
\end{equation}
where $A$ is the hyperfine coupling constant between the itinerant
electrons and the probe nuclei and $B$ is hyperfine coupling of the
spin liquid to the nuclei. Hence, if $A\ne B$ then $K_s$ will not be
proportional to $\chi$. Systematic differences between temperature
dependence of $K_s$ and that of $\chi$ are indeed found in the heavy
fermion materials.\cite{Yang} In the high temperature ($T>T_\nmr$) regime $K_s\propto\chi$ in \Br (Refs.
\onlinecite{mayaffre,kawamoto}), \NCS (Ref. \onlinecite{miyagawa})
or \CN (Ref. \onlinecite{kawamoto:kcn32006,shimizu-prb}).  We are
not aware of any reports of $K_s$ in either \Cl or \dBrn, or data
for $K_s$ at high temperatures in (TMTSF)$_2$ClO$_4$. There is only a single band that plays an important role in the organics, so one might reasonably argue that $A=B$. Nevertheless the comparison of $K_s$ and $\chi$ does not force one to consider a two fluid model and we now move on to discuss another possible  explanation of the data in Fig. \ref{fig:scaling}.

\subsection{Spin fluctuations}

Moriya's\cite{Moriya} theory of spin fluctuations in nearly antiferromagnetic metals, particularly in the phenomenological form pioneered by Millis, Monien, and Pines,\cite{MMP} has been shown to give good agreement with NMR experiments in the cuprates,\cite{MMP,Moriya} heavy fermion materials,\cite{CurroMRS,Moriya} and the organics.\cite{Eddy1}
Therefore,  it is natural to ask whether this can explain the similarities between the rescaled data for the various materials in Fig. \ref{fig:scaling}.
The simplest
assumption\cite{Moriya,MMP} for the temperature dependence of the
spin correlation length is
$\xi(T)/\xi(T_{x})=\sqrt{2T_{x}/(T+T_{x})}$. Here the temperature
dependence of the spin fluctuations is controlled by a single
parameter, $T_{x}$, thus one finds\cite{Moriya,MMP} that, in the
limit of strong spin fluctuations,
\begin{equation}
\frac{1}{T_1T} = \frac{C^2/(T_1T)_0} {(T/T_x+1)^2+4\pi^2C(T/T_x+1)}, \label{eqn:unscaled}
\end{equation}
where $C=2[\xi(T_x)/a]^2$, $a$ is the lattice spacing,
and $1/(T_{1}T)_{0}$ is a material specific constant. For $C\gg(T/T_x)+1$, i.e., if the magnetic correlation length is large and the temperature is low, one can `rescale' this to $T_\textrm{NMR}$  so that material specific
part cancels and we find that
\begin{eqnarray}
\frac{T_1T}{(T_1T)_{\nmr}} =
\frac{T_\nmr}{T_\nmr+T_x}\left(\frac{T}{T_\nmr}\right)%\nonumber\\&&
+ \frac{T_x}{T_\nmr+T_x}.\label{scaling}
\end{eqnarray}
Thus the spin fluctuation model predicts that, for nearly
antiferromagnetic metals, $T_1T/(T_1T)_{\nmr}$ is linear in
$T/T_\nmr$. This is in good agreement with the data for the metallic materials in Fig  \ref{fig:scaling}a.
Further, the spin
fluctuation theory for a nearly antiferromagnetic metal constrains
the gradient of the data to lie between 0 ($T_\nmr=0$, for a material
with neither a pseudogap nor superconductivity) and 1 ($T_{x}=0$, for
material with no spin fluctuations), cf. Eq. (\ref{scaling}).
Indeed, we find that $T_\nmr\sim T_{x}$ for all of the metallic organic charge transfer salts.
This last result is an important conclusion, as it suggests that the
pseudogap energy scale is set by that of the spin fluctuations.

Note that in (TMTSF)$_2$ClO$_4$ the maximum in the $1/T_1T$ occurs at the superconducting critical temperature, i.e., $T_c=T_\nmr\approx T_x$. This suggests that spin fluctuations may determine the superconducting critical temperature and thus the spin fluctuations may mediate the superconductivity in  (TMTSF)$_2$ClO$_4$. This is consistent with the unconventional superconducting state observed in this material.\cite{q1D}

For materials that order antiferromagnetically, one expects, within mean field theory (whence the correlation length critical exponent $\nu=1/2$), that $\xi(T)/\xi_0=\sqrt{2T_{N}/(T-T_{N})}$, where $T_N$ is the N\'eel temperature. Thus
\begin{equation}
\frac{1}{T_1T} = \frac{C_0^2/(T_1T)_0} {(T/T_N-1)^2+4\pi^2C_0(T/T_N-1)}, \label{eqn:unscaledAFM}
\end{equation}
where $C_0=2[\xi_0/a]^2$ and $a$ is the lattice spacing.
Rescaling we find that,
\begin{eqnarray}
\frac{T_1T}{(T_1T)_{\nmr}} =
\frac{T_\nmr}{T_\nmr-T_N}\left(\frac{T}{T_\nmr}\right)%\nonumber\\&&
- \frac{T_N}{T_\nmr-T_N}.\label{scalingAFM}
\end{eqnarray}
Note that the maximum in $1/T_1T$ does not occur at $T_N$ due to short range correlations. Rather, $1/T_1T$ has a infinite tangent at $T_N$, as does the bulk susceptibility.\cite{Fisher} Therefore on very general grounds\cite{Fisher} one expects $T_N<T_\nmr$. A sharp decline in $1/T_1T$ below $T_\nmr$ is observed in the data for \dBr (Ref. \onlinecite{mayaffre}) and \Cl (Ref. \onlinecite{miyagawa-kCl}), consistent with these expectations. Thus Eq. (\ref{scalingAFM}) provides an excellent description of the data for \dBr and \Cl (shown in Fig. \ref{fig:scaling}b). Note in particular that  Eq. (\ref{scalingAFM}), correctly, predicts a steep gradient and a negative intercept in agreement with the data, and in contrast to  Eq. (\ref{scaling}) and the data for the non-magnetic salts.

The rescaled NMR relaxation rate in metallic \CN (data at 0.35 and 0.4~GPa) are remarkable
similar to those in \Br and \NCS which are much more weakly
frustrated. This can be understood because $1/T_{1}T$ depends only
the the integral over the Brillouin zone of the dynamic susceptibility.
The spin fluctuation theory, which gives a good description of all of
the data, only assumes that the dynamic susceptibility has a peak
\emph{somewhere} in the Brillouin zone and that the peak is away from the origin. The predicted scaled
$T_{1}T$ is independent of the location of the peak. Thus this data
does not indicate which $\bf q$ has the strongest magnetic
fluctuations, but does show that the magnetic fluctuations are just as strong
in metallic \CN as they are in the more weakly frustrated $\kappa$-(ET)$_2X$ salts.

The rescaled NMR relaxation rate in insulating \CN (data at 0 and 0.3~GPa) is very different from that in the insulating phases of \dBr and \Cln. Instead the data for the insulating phase of \CN is well described by Eq. (\ref{scaling}), which describes the data in the metallic materials considered above. Thus it appears that the central difference between the scaled NMR relaxation rates is not whether the material is insulating or metallic but whether the material orders magnetically or not. We note that both data sets fit remarkably well to the prediction of Eq. (\ref{scaling}) for $T_x=1.76 T_\nmr$ in the insulating phases of \CNn. This brings to mind the   weak coupling BCS form formula for the pairing temperature in a superconductor. However, at this stage we have no evidence that this is any more than a numerical coincidence.

%It is interesting to observe that both data sets fit remarkably well to the prediction of Eq. (\ref{scaling}) for $T_x=1.76 T_\nmr$ in the insulating phases of \CNn. This is consistent with a gap of magnitude $T_x$ opening below $T=T_\nmr$ that has a week coupling BCS form. Is this a sign the spin fluctuations cause open a gap in \CN at low temperatures? Certainly the Knight shift decreases rapidly below $T_\nmr\sim10$ K.\cite{kawamoto:kcn32006} However, both the maximum in $1/T_1T$ and the decrease in $K_s$ are much broader than one would expect for a phase transition. It is also interesting to note that a broad hump is observed in the heat capactity below $\sim$8 K (Ref. \onlinecite{Yamashita}), i.e. just below $T_\nmr$. It is not clear at present whether these two features are related. However, if they are, the idea that $T_\nmr$ is determined by spin fluctuations contradicts the proposal of Lee \etal\cite{Lee} that an instability in the spinon liquid is caused by the effective $U(1)$ gauge field.

%\subsubsection{(TMTSF)$_2$ClO$_4$}

\section{Pseudogap and coherence}\label{sect:coh}

\begin{figure}
\epsfig{file=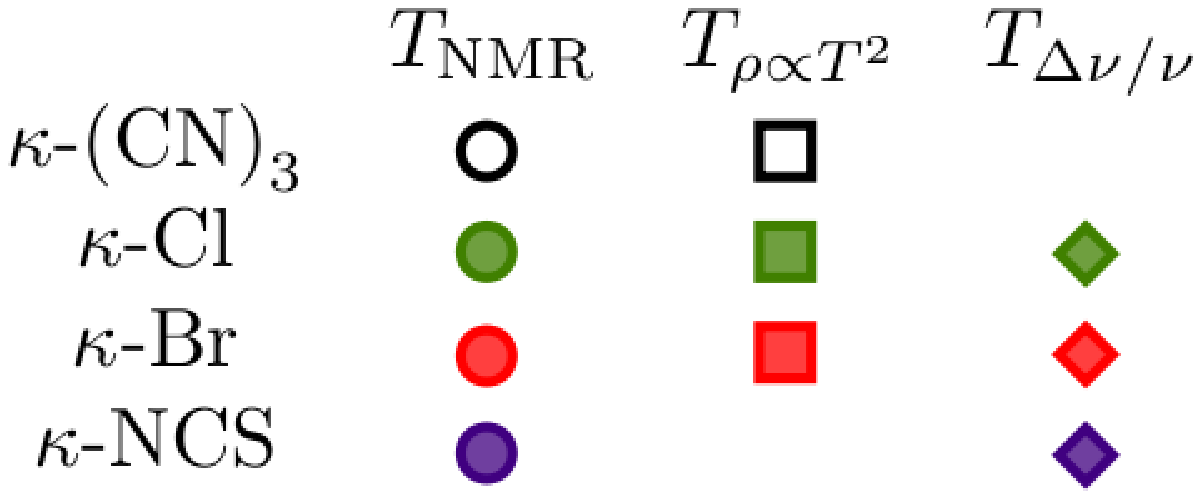, width=5cm}
\epsfig{file=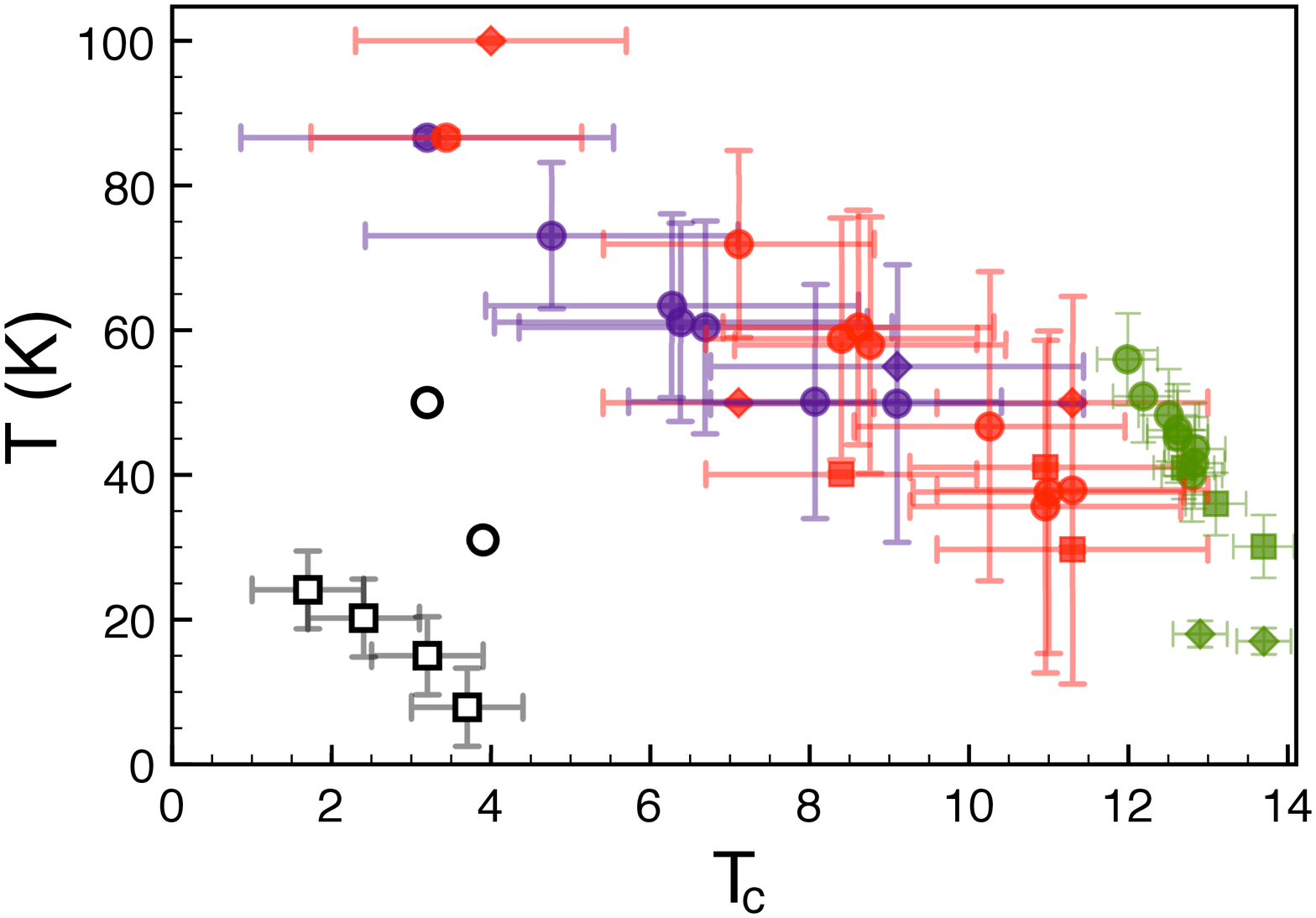, width=8cm}
\caption{[Color online]
Variation of experimental energy scales in $\kappa$-(ET)$_2X$ 
with hydrostatic pressure. As, in each material, the superconducting transition temperature decreases monotonically with the applied pressure the
superconducting transition temperature serves to parameterise  the
proximity to the Mott transition (high $T_{c}$s closest to the Mott
transition).
The fact that that $T_{\nmr}$,
$T_{\res}$, and $T_{\us}$ show such similar behaviours in \Cln, \Brn,
and \NCS is a success for the chemical pressure hypothesis. The chemical pressure hypothesis is seen to fail
dramatically in the case of \CNn,  which shows a markedly
different behaviour to the other salts. Further in \CN $T_{\nmr}$
does not coincide with $T_{\res}$.  We stress that the large error
bars in the figure result predominately from the errors measuring the
hydrostatic pressure which appears parametrically in comparing
multiple different experiments.
Data is taken from Refs. \onlinecite{lefebvre,fournier,frikach,schirber,caulfield,
kurosaki,limelette,mayaffre,kawamoto,kawamoto:kcn32006,strack}.} \label{fig:phase}
\end{figure}

We now turn our attention to how the pseudogap energy scale,
$T_{\nmr}$, is related to the other energy scales in the $\kappa$-phase organic
charge transfer salts. Fig. \ref{fig:phase} shows that, for the
weakly frustrated salts \{\Cln, \Brn, and \NCSn\} $T_{\nmr}$
coincides with the temperature below which the resistivity varies
quadratically, $T_{\res}$, and the temperature at which a broad, deep
minimum is observed in the ultrasonic velocity, $T_{\us}$. Further,
when these quantities are plotted against the superconducting
transition temperature, $T_{c}$, all three materials show the same
trend.
%This correlation does not necessarily
%demonstrate a causal link between $T_{c}$ and
%$T_{\nmr}$, $T_{\res}$, and $T_{\us}$. Nevertheless,

Fig. \ref{fig:phase} represents a
significant quantitative success for the chemical pressure hypothesis, which
holds that the major effect of changing the anions is to alter size
of the unit cell and therefore control the strength of the electronic
correlations
and which underpins a great deal of the thinking on the organic charge
transfer salts. 

However, the chemical pressure hypothesis clearly
fails for \CNn, which does not share the same scaling of
temperature scales as the other salts investigated. Further, $T_\nmr$
is clearly rather different from $T_\res$ in \CNn. This suggests that
the metallic state of \CN is rather different from the metallic
states of the more weakly frustrated $\kappa$-phase organic charge transfer salts, consistent with the idea that the spin fluctuations in this material differ from those in its less frustrated cousins in important ways.\cite{group-theory,d+id}
%This is direct evidence that the  spin fluctuations in a metal near a
%spin liquid are significantly different from those near in a nearly
%antiferromagnetic metal.
Nevertheless the observation of a quadratic temperature dependence of the resistivity at temperatures slightly above $T_c$ shows that charge transport is coherent in the metallic state of \CN under pressure.

The observation that the crossover to a quadratic temperature dependence of the resistivity approximately coincides
with the minimum in the ultrasonic attenuation has been
understood\cite{Merino,limelette} in terms of dynamical mean field
theory (DMFT). DMFT predicts a crossover from a Fermi liquid at low
temperatures to  a `bad metal' (characterised by the absence of
quasiparticles and the Drude peak, and a resistivity that exceeds the
Mott-Ioffe-Regal limit) as the temperature is increased above a
coherence temperature, $T_\textrm{coh}$. Thus it is believed that
$T_\textrm{coh}\simeq T_{\res}\simeq T_{\us}$. However, while DMFT gives an
adequate description of the nuclear spin relaxation rate above
$T_\textrm{coh}\simeq T_\nmr$, below $T_\coh$ DMFT predicts a
Fermi liquid and thus a constant $1/T_1T$; this is clearly \emph{not} what is
observed.\cite{Eddy1} This shows that DMFT does not capture all of
the relevant physics below $T_\coh$.\cite{Eddy2} It is also interesting to note that DMFT does describe the observed
behaviour of probes of charge degrees of freedom, such as the
resistivity, and only fails for probes of the spin degrees of
freedom, such as $1/T_1T$. As DMFT is a purely local theory
a reasonable hypothesis it that the relevant physics, not described by DMFT, involves short-range spin correlations. Further, the enhanced Korringa ratios observed\cite{Eddy1} at temperatures slightly above $T_c$ suggest that antiferromagnetic correlations remain important at in the coherent transport regime.\cite{Doniach,Eddy2}

The spin degrees of freedom in the pseudogap regime of the
$\kappa$-phase organics behave in much the same way as the spin
degrees of freedom in the pseudogap regime of the underdoped
cuprates. There has been significant debate as to whether, and if so
how, the pseudogap in the cuprates is related to the other exotic
phenomena seen in the normal state, such as the linear temperature dependence of the resistivity. In this context it is
interesting to note that in the region of the phase diagram of the
organics where the pseudogap is found, the resistivity varies
quadratically with temperature\cite{limelette} and the magnitude of the coefficient
of the quadratic term in the resistivity is as expected from Fermi
liquid theory given the observed effective mass.\cite{JFP} Further,
clear evidence of quasiparticles is seen via quantum oscillation
experiments.\cite{singleton}  Therefore,  in \Brn, \dBrn, and \NCS the loss of spectral weight in the pseudogap is not associated with non-Fermi--liquid behaviour.

An important difference between the
cuprates and the organics is that the organics are half filled
whereas the cuprates are more strongly correlated, doped systems. In
this context it is worth noting that a linear resistivity has
recently been reported in an organic charge transfer salt with an
anion layer that has a lattice constant that is incommensurate with the
lattice constant of the organic layer.\cite{Taniguchi} The authors
argued that this non-stoichiometric organic charge transfer salt is
effectively doped away from half filling.

\section{High pressures}\label{sect:highP}

An important question in understanding the phenomenology of the
$\kappa$-(ET)$_2X$ salts is: do the two energy scales, $T_\coh$ and
$T_\nmr$ remain equal as pressure is increased and we move further
from the Mott transition? This question is difficult to answer at
present because there is little experimental data for high pressures
(including high chemical pressure, i.e., materials with low
$T_c$'s). There is however tantalising evidence that something
rather interesting happens to the superconducting state at high
pressures.\cite{Tc-penetration} In particular while the materials
near the Mott transition have a superfluid stiffness, $n_s$, within a factor
of two or so of the prediction of BCS theory, at high pressures the
superfluid penetration depth, $\lambda$, increases as $T_c\propto1/\lambda^3$
(Ref. \onlinecite{Pratt}). With some materials having  superfluid
stiffnesses ($n_s\propto1/\lambda^2$) that are an order of magnitude smaller than the prediction of BCS
theory.\cite{Pratt,Tc-penetration} 

We are only aware of one NMR
experiment at high pressures in these materials. Ref.
\onlinecite{mayaffre} reports data for \Br at 3 kbar (which leads to
$T_c\simeq3.8$ K)\cite{schirber} and 4 kbar ($T_c\simeq1.4$ K).  At
these pressures strong spin fluctuations are not observed, and the
$1/T_1T$ looks quite conventional. Yet for the 3 kbar data there is
small but noticeable decrease in $1/T_1T$ below  $\sim20$ K. Is this
the last vestige of the pseudogap? If so, it suggests that at high
pressures the pseudogap and the coherent intralayer transport energy scales are
different. Either way more experiments are clearly required to
understand where the pseudogap vanishes.

\begin{figure}%[h]
\epsfig{file=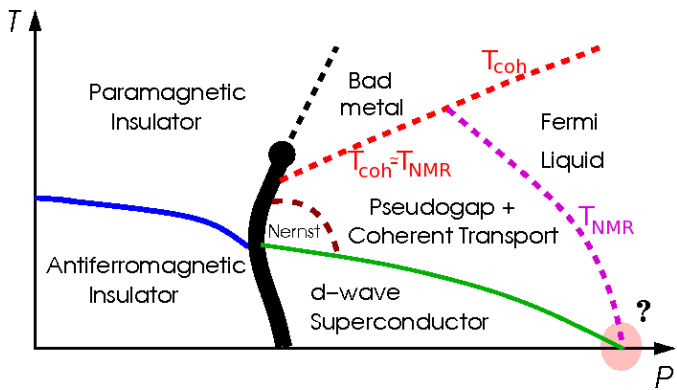,scale=0.5} \caption{[Color online]
Schematic phase diagram for weakly frustrated \cation X as a
function of temperature and pressure. Thin solid lines represent
second order phase transitions, the thick solid line is the first
order transition line which ends at a critical point shown as a
filled circle, and dashed lines indicate crossovers. The pseudogap
regime is much more complicated than a renormalized Fermi liquid
that has been previously thought to characterize the paramagnetic
metallic phase at low temperatures. It shows a coherent transport
character with long lived quasiparticles, marked by $T^2$
resistivity behavior\cite{limelette} with the coefficient of the
quadratic term as expected from Fermi liquid theory given the observed
effective mass,\cite{JFP} and magnetic quantum
oscillations.\cite{singleton} But, a loss of
spectral weight is clearly observed in the NMR data. There are not
sufficient data at this moment to determine what happens to the
pseudogap regime at high pressures; this uncertainty is represented
by the shaded area with the question mark.}\label{fig:new_phase}
\end{figure}

On the basis of the above discussion we propose that a number of new
features should be included in the phase diagram of these materials,
which we sketch in Fig. \ref{fig:new_phase}. We stress that this
phase diagram is relevant to the weakly frustrated materials for
which the chemical pressure hypothesis holds, and therefore does not
include \CN which would necessitate an additional axis to
include the effects of frustration. We have included the Nernst
region where Nam \etal\cite{Nam} have observed a large Nernst effect
above $T_c$ in \Br at ambient pressure, which they find to be absent
in \NCS at ambient pressure. Note that we have drawn
the $T_\nmr$ and the $T_c$ lines both supressed to zero at the same
pressure. This is deliberately provocative. As we have stressed above there is
insufficient experimental data to determine the relative order in
which the superconductivity and the pseudogap disappear as pressure
is increased.

The issue of where the superconductivity and pseudogap vanish is related to an ongoing debate in the cuprates. A recent review of a wide range of experimental data in a wide variety of cuprates suggested that the pseudogap and superconductivity both vanish at the same critical doping.\cite{Hufner}
Similarly, more detailed
experimental and theoretical studies of the organics in the vicinity of the pressure above which superconductivity vanishes
might give important insights into how the pseudogap is related to
superconductivity. The possibility of a quantum critical point
somewhere in the vicinity of the pressure where the superconducting
critical temperature goes to zero may have important consequences
for the observation that the materials with the lowest
superconducting critical temperatures have extremely small
superfluid stiffnesses and are very different from BCS
superconductors.\cite{Pratt,Tc-penetration}

Finally, we sketch an explanation of the observed phase diagram (Fig.
\ref{fig:new_phase}). We have noted above that DMFT describes the
competition between the insulating, bad metal, and Fermi liquid
phases.\cite{Merino,limelette} But, DMFT fails to predict either the
pseudogap or the unconventional superconductivity, which suggests
that these effects involve non-local physics. However, it has
recently been argued that the resonating valence bond (RVB) theory
can describe the superconducting state of the quasi-two-dimensional organic charge transfer
salts,\cite{organicsRVB,d+id} this theory also predicts a pseudogap
with the approximately the right energy scale. Therefore, we propose
that DMFT captures the crossover from incoherent to coherent charge
transport, but is insufficient to describe the behaviour in the
coherent regime because short-range spin correlations play a significant role
here. The simplest theory that can capture the low temperature
physics of is the RVB theory. However, the RVB theory does not
capture the loss of coherence as the temperature is raised, which we
argue leads to a cut off of the pseudogap phenomena. If this
speculation is correct, the challenge is then to produce a single
theory capable of describing all of the physics, including the large
Nernst effect above $T_c$ in \Br (Ref. \onlinecite{Nam}), which neither the DMFT nor the RVB
theory predicts. Therefore cellular DMFT calculations, which can describe both short-range spin fluctuations and the loss of intralayer coherence, may have an important role to play in understanding the organic superconductors.

\section{Conclusions}\label{sect:conc}

We have argued that a two-fluid description is not required for the
organic charge transfer salts. It is interesting to speculate why
this is. One possibility is that the difference between the organic
salts discussed above and many other strongly correlated
superconductors is that the organics are stoichiometric, while, for
example, the metallic cuprates are doped systems and the heavy
fermions show a subtle hybridisation between almost localised states
and conduction electrons. Therefore it would be interesting to
measure the nuclear relaxation rate in the recently discovered
non-stoichometric organic superconductors.

We have seen that there are strong similarities in the rescaled spin lattice relaxation across the non-magnetic organic charge transfer salts. These similarities can be understood in terms of spin fluctuations. Further, our analysis suggests that the energy scale for the spin fluctuations may set the energy scale for the pseudogap in the organic charge transfer salts.

In the weakly frustrated metals [all the metals studied bar \CNn]
the pseudogap opens at the same temperature as coherence emerges in
the (intralayer) transport. We argued that this is because spin
correlations, which are responsible for the pseudogap, are cut off
by the loss of intralayer coherence at high temperatures. In these
weakly frustrated materials the data can be compared across
materials quite reliably, consistent with the chemical pressure
hypothesis. In contrast the metallic state of  \CNn, which is highly
frustrated, was shown to be rather different from those of the other
materials.

Finally, we have also shown that in (TMTSF)$_2$ClO$_4$ the characteristic temperature scale of the spin fluctuations is the same as $T_c$. This suggests that spin fluctuations may mediate the superconductivity in the Bechgaard salts.

\section*{Acknowledgements}

We would like to thank S. Brown, K. Kanoda, J. Merino, and H. Monien
for useful conversations.  This work was supported in part under
Australian Research Council's Discovery Projects funding scheme
(project DP0557532). B.J.P. is supported  by the Australian Research Council under the Queen
Elizabeth II scheme (project DP0878523). R.H.M.  is supported  by the Australian Research Council under the APF scheme (project DP0877875).

\end{document}